# Technology utilization patterns and business growth in Small/Medium Enterprises

Chaitanya Dhareshwar

Date: 1st June 2018

## Abstract

Technology was created to support and grow Business. Modern business that uses technology efficiently, grows at a phenomenal rate (Statista.com, 2018). The assumption therefore is that businesses that utilize insufficient technology, or use technology inefficiently, experience reduced growth and possibly, business decline.

Technological development holds great significance in most industries particularly in wastage reduction, process optimization and consequently bottom-line revenue enhancement and price-leadership. We've seen revolutionary technological development during the 20th century / 21st century thus far, (Ivanovic et al, 2015) and it's led to drastic growth in fields like communication, computer science, monitoring of operations, remote working, high performance analytics and many more. Some fields have even come into existence purely due to technology.

Technological equipment cannot compensate for the skills, knowledge or creativity of human employees. However, expertise of the average employee can be greatly enhanced using intelligent software. Use of such equipment decreases the need for unskilled and semi-skilled workers – but can exponentially increase speed of performance for skilled workers. Innovations are key defining criteria for competitive differentiation – but some of these can be easily copied, which basically means that innovation and improvement are continuous processes. Process standardization comes through in a big way when technological solutions are applied in the work. It regulates/optimizes the number of employees needed, power consumption, potentially reduces wastage, drastically improves hygiene process (where relevant). The natural outcome is greater process efficiency and cost efficiency.

# Introduction

Technological implementation can significantly enhance business bottom-line revenues through process improvement, wastage reduction, automation of high-risk activities, etc. Some work has already been done in this area, and it would be remiss to not cite the same ref (Lindh, 2006). Although – in contrast to this paper – they have focused upon the benefits of good technology application and utilization. We have seen the employment of information technology among industrial companies resulting in improvements and cost-reduction (Deeter-Schmelz, 2002, Pires and Aisbett, 2003). Some technology solutions enable geographically dispersed users to share data and messages, even instantly, to any number of recipients (Claycomb et al., 2004, Deeter-Schmelz 2002, Reunis et al., 2005, Öhrwall Rönnbäck, 2002).

Using for example Email or other Electronic Data interchange sends information flows entirely through information technology, such as the e-Commerce and Logistics, or Payment information (Angeles and Nath, 2000). Integrating EDI measurably increases efficiency and permits businesses to save time, perhaps to reduce costs. (Hill and Scudder, 2002, Laage-Hellman, 1989).

In recent years, many researchers have focused upon the development of information technology, and its beneficial effect upon modern business (Chatfield and Yetton, 2000, Leek et al., 2003, Vlosky and Smith, 1993, Wilson and Vlosky, 1998). Some claim that information technology increases business efficiency other researchers explore the negative impacts of technology.

This paper presents the outcome of one statistically relevant study out of several, for the purpose of peer review and discussion. The expectation, from showcasing this specific study, is to draw sufficient conclusion that quantifies and qualifies the value of technology use in business today. We try to realize what advantages technology brings (in ROI terms), and whether succession planning for technology itself, holds any value.

Note: Target scope optimizations – IT-based or otherwise tech focused companies are excluded from the scope – although the type of technology used in these for 'operations' and 'business enhancement' are different, there's insufficient statistical evidence of reduced growth / business decline primarily due to technology reasons. Other reasons such as early-for-market, poor implementation, bad go-to-market strategy were found to be more relevant there (but will not be covered in this analysis).

Presented here is an empirical analysis of 19 Micro, Small and Medium Enterprises (MSMEs) across the verticals of Healthcare, FMCG, Logistics and Hospitality/Tourism. (MSME: http://ec.europa.eu/growth/smes/business-friendly-environment/sme-definition_en)

Table 1: Verticals of the companies included are as follows

| Vertical | Approached | Included in Analysis |
|---|---|---|
| **FMCG** | 18 | 7 |
| **Healthcare** | 22 | 5 |
| **Logistics** | 5 | 3 |
| **Hospitality/Tourism** | 8 | 4 |
| **Other service businesses** | 15 | 0[*] |
| **Other product businesses** | 17 | 0[*] |

* = Those not included in the analysis either chose not to participate, or provided insufficient statistical evidence to be usable for the study.

Table 2: Organizational structure and reach of these organizations fell in the ranges noted below.

| | Min | Max | Median | Average |
|---|---|---|---|---|
| **Employee Count** | 10 | 300 | 220 | 211 |
| **Customers** | 1[*] | 50,000 | 4,800 | 10,886 |
| **Board size** | 1[#] | 6 | 3 | 3 |

Numbers are rounded off, and thus approximate within +/- 2%

* = This company primarily serves just one single customer and refused other business opportunities

# = This is a one-person company

Countries included in the analysis: Bangladesh, El Salvador, Georgia, India, Indonesia, Kenya, Philippines, Zambia. Choice of countries is detailed in the Method > 'Selection of Target Audience' section.

## Method

Results presented are derived from statistical analysis of the outcome of a questionnaire circulated to the companies "Included in Analysis" as noted in the introduction. Aggregate results are provided in the Appendices.

The research consisted of steps noted here:

1. Target Audience definition: The choice of target audience was based upon two factors.
   a. SMEs comprise 99% of enterprises in the EU (ref: http://ec.europa.eu/eurostat/statistics-explained/index.php/Statistics_on_small_and_medium-sized_enterprises).
   b. The EU is a large aggregate of disparate countries and makes a statistically acceptable representation of World population (assumption).
2. Selection of Target audience: This involved a review of MSME based upon certain criteria including Country, Population thereof, % of MSMEs, % Employment, and value added to the economy (Ref: https://www.smefinanceforum.org/data-

sites/msme-country-indicators). Some selectivity in terms of countries was required however, thus these criteria were added:
   a. Countries that have a significant IT-driven or tech-focused MSME presence were eliminated, as these organizations exclude our target scope.
   b. High-revenue countries that disproportionately increased MSME revenues (usually developed countries, revenues more than $10M) were eliminated as outliers.
   c. MSMEs that worked with marginal improvement situation and did not result in long term value creation, or companies that did not control a significant part of the 'value chain' (e.g.: drop shipping, pass-through merchants, affiliate marketing companies) were excluded, due to poor tech-optimization possibilities.
3. Finally, 85 choice companies were identified in terms of:
   a. Email accessibility (company's domain supports incoming email with valid MX record and receives on port #25)
   b. Presence of a Chief Technology Officer / IT Head / VP IT / GM IT / IT Manager role in the organization (identified clearly via Linkedin)
   c. Company appears to be active in regional registry. Where a regional / country corporate registry was not clearly identifiable, the company was excluded.
   d. Intended target person in the company was present on LinkedIn and held a top-level role in the organization equivalent of a C-level or were Directors in the company.
4. Outreach methodology: Individual invites and requests were sent to the CXOs of these companies. Responses were varied.
   a. 40% did not respond at all, or did not respond positively.
   b. Of the ~62% that did respond, some did not provide usable data (either left the form half-way, or did not provide data within predetermined acceptable bounds)
   c. Final list of valid respondents with sufficient data to analyse = 19.
5. Data collection methodology: One Google Sheet form was prepared with questions (as further noted in the Appendices), and supplied to willing participants of the survey. The result was downloaded as an Excel sheet.
Note: Names of companies and names of the respondents were optional in keeping with anonymity requirements of certain regions (specifically the GDPR for some European countries). Of the respondents that filled some part of the form, ~85% (45 respondents) chose not to identify themselves. Name information is hence excluded in furnished data.
6. Analysis methods: Data was analysed using Ms Excel, observations made visually and graphically.

## Observations

1. Businesses that have enabled tech have seen higher business value through tech
   a. Companies that were tech based from inception have rated the business value seen as 7.8 (out of 10), but this was only 6.3 for companies that were not 'built upon' tech.
   b. Companies that started leveraging tech before 2010 felt business value from the same was 7.6 (out of 10), while those after 2010 (late adopters) felt 5.2 was what they experienced

   Thus, one can reasonably infer that having more technology-based leverage, ideally from inception of the company, can significantly change the business value from that tech.

2. MSMEs spend an average of ~USD 67,000 when moving from a non-tech based organizational process to a tech-based process. This expense didn't exist in companies that were tech-based at inception, because:
   a. A majority of processes were automated using existing tech best-practices
   b. Little specialized software/hardware was needed to serve the objectives
3. Two companies chose not to get websites / mobile apps done, because they felt it would not add to visibility or organizational growth. One of them has recently started tech initiatives, but will not be building a website as they feel it would be a frivolous effort.
4. Other than the two companies that chose not to get websites done, one company that focused upon a single customer realized higher visibility but chose to not accept more leads/customers. Yet another company didn't see a sizeable increase of value in bookings (as was their expectation from creating a website) – the cause of this is unclear but appears to be linked with marketing efforts (they admit to zero marketing budget).
5. Companies that implemented Process Automation, felt that it added enough value to justify a rating of 8.13 (of 10). 3 companies (one of which is single-customer focused) did not engage with process automation because they felt the human way was sufficient. There was no objective measurement made for this by the companies, thus no justification was possible.
6. Except for one company, all experienced a waste reduction in their operating processes. The company that didn't experience waste reduction also did not have a digital waste measurement mechanism in place – whereas the others did either as part of their ERP or other inventory management mechanism.
7. All 19 businesses have working payment gateways in place, and agreed that payment gateways resulted in faster conversions.
8. All but two companies felt they experienced much better customer retention, and the same companies also agreed upon an increased conversion ratio. The exceptions are: the company that had just one customer, and the other respondent didn't answer these two questions.

9. All respondents that had started without a technological base and later moved their business processes to tech felt that the ROI for this move was roughly equal to the investment. Those that had started with a tech base could not quantify ROI quite so well, the main reason given when asked verbally was: "It's a core part of business and thus inseparable from our bottom line results.". This was unanimous except for one company that chose to quantify first year ROI (i.e. at inception) as USD 11,000.

10. ROI from technology initiatives in the past year (i.e. 2017) for most companies is directly proportional to the time spent in a tech-based environment.

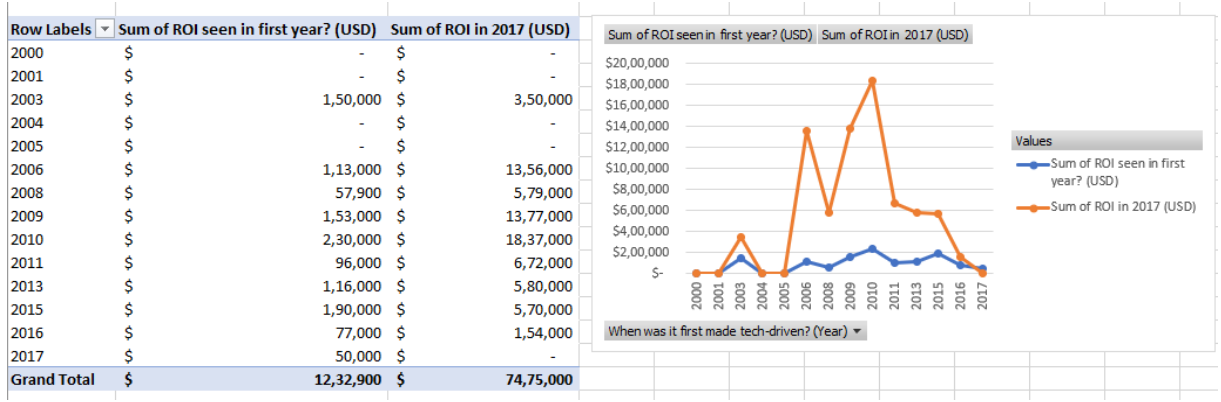

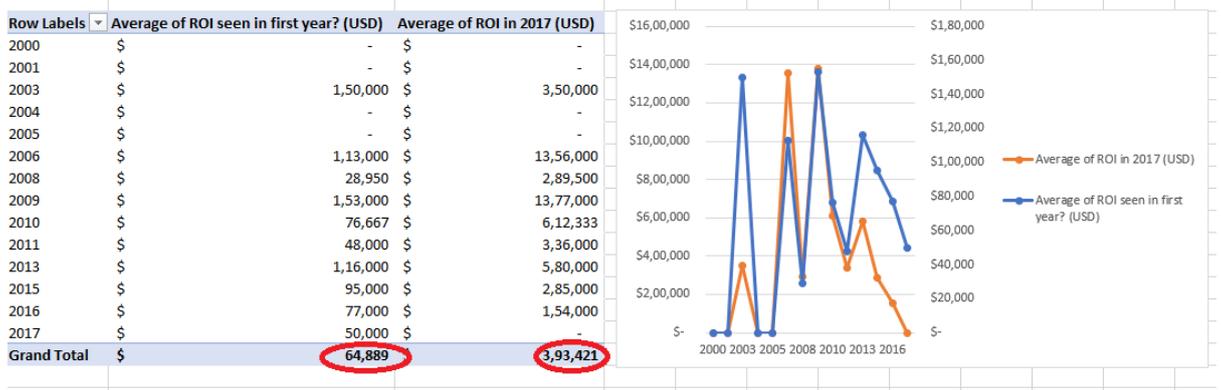

In some cases, the respondents have allotted their entire business revenue to this field, with the logic that tech has now become a core part of their business.

11. The graph noted above shows a curve that seems to indicate that technology ventures have an optimal outcome duration of 5-12 years. Recent ROIs for companies in this range are several times the first-year ROI.

12. One company that had started tech initiatives as recently as 2017 has seen some positive results, but they've put this figure under 'first year ROI' and left 'ROI in past year' blank, this figure is ignored in the ROI calculations.

13. A one-person company with a small employee base is able to handle ~5000 customers with ease thanks to the use of time saving technology.

14. Businesses that have moved on tech activities recently (past 3 years) have seen a slight dip in revenue (due to process restrictions / standardization) when the changes were implemented, but expect to make up the gap rapidly

15. Organizations despite a low employee count and low customer count were able to generate significant ROI.

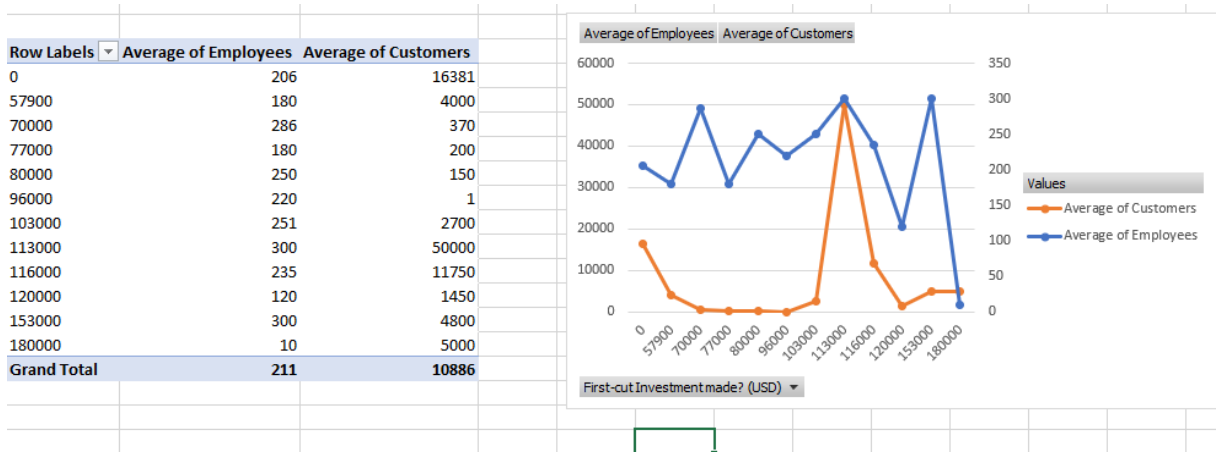

| Row Labels | Average of Employees | Average of Customers |
|---|---|---|
| 0 | 206 | 16381 |
| 57900 | 180 | 4000 |
| 70000 | 286 | 370 |
| 77000 | 180 | 200 |
| 80000 | 250 | 150 |
| 96000 | 220 | 1 |
| 103000 | 251 | 2700 |
| 113000 | 300 | 50000 |
| 116000 | 235 | 11750 |
| 120000 | 120 | 1450 |
| 153000 | 300 | 4800 |
| 180000 | 10 | 5000 |
| Grand Total | 211 | 10886 |

This may also be looked at the other way, in that organizations with fewer customers and fewer employees spent more on tech; and thus, experienced good ROI.

## Conclusion

1. Company-wide technology initiatives take an average of 1-2 years before they start paying for themselves and generate growing returns over a period of time.
2. Investing in tech can result in greater returns irrespective of number of customers or number of employees – fundamentally it comes down to whether the tech in use is able to perform the tasks intended and drive bottom-line impacting outcomes.
3. Potentially, companies with a size of 2-5 people if well equipped with the correct technology can handle a huge customer base without difficulty. (Lean companies)
4. For the average company that is financially sound, an investment in technology today can potentially generate more than 100% returns in a span of 3 years.
5. Investment in this sequence of importance:
    I. Payment gateway
    II. Process automation
    III. Website/mobile app

    Appears to generate good results.

6. Since older technology needs replacement from time to time, and returns start to diminish after 7-12 years, it is most important to create a technology succession plan where upgrade and replacement paths for all technology initiatives is clearly defined.
7. Obsolescence is a real risk - one must note that the average duration is between 3~4 years. This represents a challenge and a consistent 3rd party technology hub may prove to be invaluable to help mitigate this risk.

## References

Up-to-date references may be obtained at https://cbd.vcio.in/2018/06/01/chaitanya-dhareshwar-2018-june-technology-utilization-patterns-and-business-growth-in-small-medium-enterprises/

| # | Customers | Employees | Board of Directors | Was it always so? | When was First-cut In place | Website / Tech spend | Rate business impact | Increased visibility | Increased leads | Increased customers | Automate processes | Rate business impact | Savings of time | Waste reduction | Faster outcomes | Payment gateways | Rate business impact | Better customer experience | Faster communication | Increased customers | ROI seen in first year | ROI in 2017 (USD) | Do you expect more | Comments |
|---|---|---|---|---|---|---|---|---|---|---|---|---|---|---|---|---|---|---|---|---|---|---|---|---|
| 1 | 5000 | 10 | 1 | N | 2003 | 180000 Y | 9 Y | Y | Y | Y | | 10 Y | N | Y | Y | | 7 - | Y | - | $ 1,50,000.00 | $ 3,50,000.00 | Y | 1. Tech expense ballooned from $5000 -> $180,000 when we got ambitious 2. ROI was proportional to investment, thus agreed to invest more 3. No discernable change in wastage. But no 'new' digital waste tracking mechanism in place. |
| 2 | 50000 | 300 | 5 | N | 2006 | 113000 Y | 7 Y | Y | Y | Y | | 9 Y | Y | Y | Y | | 8 Y | Y | Y | $ 1,13,000.00 | $ 13,56,000.00 | Y | |
| 3 | 3000 | 177 | 2 | Y | 2005 | 0 Y | 8 Y | Y | Y | Y | | 8 Y | Y | Y | Y | | 10 Y | Y | Y | $ - | $ - | Y | |
| 4 | 200 | 180 | 3 | N | 2016 | 77000 Y | 6 Y | Y | Y | N | - | - | - | - | Y | | 9 Y | Y | Y | $ 77,000.00 | $ 1,54,000.00 | Y | |
| 5 | 1 | 220 | 4 | N | 2011 | 96000 Y | 6 Y | N | N | N | - | - | - | - | Y | | 10 Y | N | Y | N | $ 96,000.00 | $ 6,72,000.00 | Y | |
| 6 | 6000 | 253 | 2 | Y | 2008 | 0 Y | 8 Y | Y | Y | Y | | 9 Y | Y | Y | Y | | 9 Y | Y | Y | $ - | $ - | Y | |
| 7 | 4800 | 300 | 2 | N | 2009 | 153000 Y | 7 Y | Y | Y | Y | | 8 Y | Y | Y | Y | | 9 Y | Y | Y | $ 1,53,000.00 | $ 13,77,000.00 | Y | |
| 8 | 5500 | 169 | 3 | N | 2010 | 116000 Y | 5 Y | Y | Y | N | - | - | - | - | Y | | 6 Y | Y | Y | $ 1,16,000.00 | $ 9,28,000.00 | Y | |
| 9 | 370 | 286 | 1 | N | 2015 | 70000 N | - | - | - | - | Y | | 6 Y | Y | Y | Y | | 7 Y | Y | Y | $ 70,000.00 | $ 2,10,000.00 | Y | |
| 10 | 18000 | 300 | 1 | N | 2013 | 116000 Y | 5 Y | Y | Y | Y | | 6 Y | Y | Y | Y | | 8 Y | Y | Y | $ 1,16,000.00 | $ 5,80,000.00 | Y | |
| 11 | 44300 | 297 | 3 | Y | 2011 | 0 Y | 3 Y | Y | Y | Y | | 8 Y | Y | Y | Y | | 7 Y | Y | Y | $ - | $ - | Y | |
| 12 | 23670 | 183 | 3 | Y | 2000 | 0 Y | 9 Y | Y | Y | Y | | 7 Y | Y | Y | Y | | 8 Y | Y | Y | $ - | $ - | Y | |
| 13 | 17000 | 250 | 4 | Y | 2001 | 0 Y | 9 Y | Y | Y | Y | | 7 Y | Y | Y | Y | | 7 Y | Y | Y | $ - | $ - | Y | |
| 14 | 3000 | 190 | 3 | Y | 2004 | 0 Y | 8 Y | Y | Y | Y | | 8 Y | Y | Y | Y | | 10 Y | Y | Y | $ - | $ - | Y | |
| 15 | 2700 | 251 | 6 | N | 2010 | 103000 Y | 4 Y | Y | Y | Y | | 9 Y | Y | Y | Y | | 10 Y | Y | Y | $ 1,03,000.00 | $ 8,24,000.00 | Y | |
| 16 | 150 | 250 | 5 | N | 2017 | 80000 N | - | - | - | - | N | - | - | - | - | Y | | 9 Y | Y | Y | $ 50,000.00 | $ - | Y | 1. Founder felt tech is not a good fit for them initially 2. Saw competing companies lean heavy on tech and grow 3x-8x 3. Took the leap in 2017, but as-yet unsure as their tech is not in place yet |
| 17 | 17700 | 93 | 3 | Y | 2010 | 0 Y | 10 Y | Y | Y | Y | | 10 Y | Y | Y | Y | | 8 Y | Y | Y | $ 11,000.00 | $ 85,000.00 | Y | 1. Started with tech in place 2. Growth was ~2x YoY, with some fluctuations in 2015-2018 3. Founder puts down growth to tech more than staff/team/idea/market |
| 18 | 4000 | 180 | 2 | N | 2008 | 57900 Y | 8 Y | Y | Y | Y | | 8 Y | Y | Y | Y | | 10 Y | Y | Y | $ 57,900.00 | $ 5,79,000.00 | Y | |
| 19 | 1450 | 120 | 4 | N | 2015 | 120000 Y | 6 Y | N | N | Y | | 9 Y | Y | Y | Y | | 9 Y | Y | Y | $ 1,20,000.00 | $ 3,60,000.00 | Y | |
| avg | 10886.37 | 211 | 3 | | | | | | | | | | | | | | | | | | | | |
| med | 4800 | 220 | 3 | | | | | | | | | | | | | | | | | | | | |
| max | 50000 | 300 | 6 | | | | | | | | | | | | | | | | | | | | |
| min | 1 | 10 | 1 | | | | | | | | | | | | | | | | | | | | |

| Row Labels | Average of ROI seen in first year? (USD) | Average of ROI in 2017 (USD) |
|---|---|---|
| 2000 | $ - | $ - |
| 2001 | $ - | $ - |
| 2003 | $ 1,50,000 | $ 3,50,000 |
| 2004 | $ - | $ - |
| 2005 | $ - | $ - |
| 2006 | $ 1,13,000 | $ 13,56,000 |
| 2008 | $ 28,950 | $ 2,89,500 |
| 2009 | $ 1,53,000 | $ 13,77,000 |
| 2010 | $ 76,667 | $ 6,12,333 |
| 2011 | $ 48,000 | $ 3,36,000 |
| 2013 | $ 1,16,000 | $ 5,80,000 |
| 2015 | $ 95,000 | $ 2,85,000 |
| 2016 | $ 77,000 | $ 1,54,000 |
| 2017 | $ 50,000 | $ - |
| **Grand Total** | **$ 64,889** | **$ 3,93,421** |

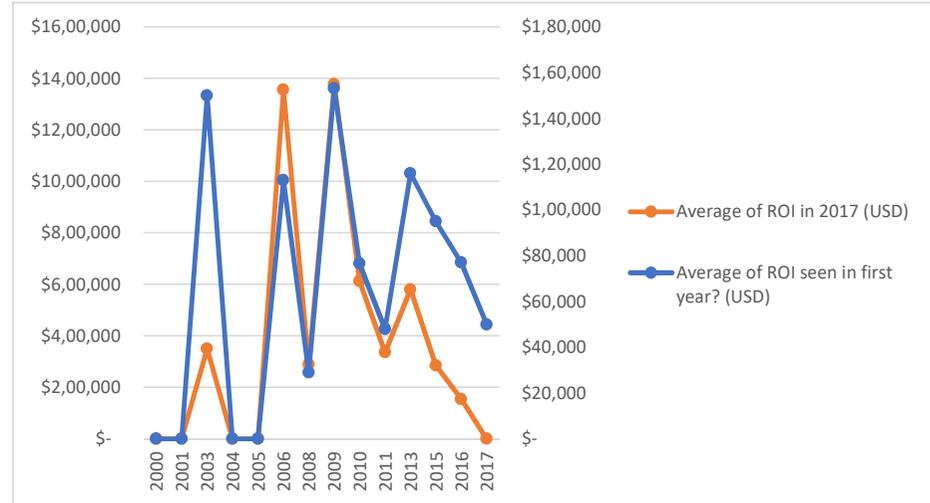

| Row Labels | Average of Employees | Average of Customers |
|---|---|---|
| 0 | 206 | 16381 |
| 57900 | 180 | 4000 |
| 70000 | 286 | 370 |
| 77000 | 180 | 200 |
| 80000 | 250 | 150 |
| 96000 | 220 | 1 |
| 103000 | 251 | 2700 |
| 113000 | 300 | 50000 |
| 116000 | 235 | 11750 |
| 120000 | 120 | 1450 |
| 153000 | 300 | 4800 |
| 180000 | 10 | 5000 |
| Grand Total | 211 | 10886 |

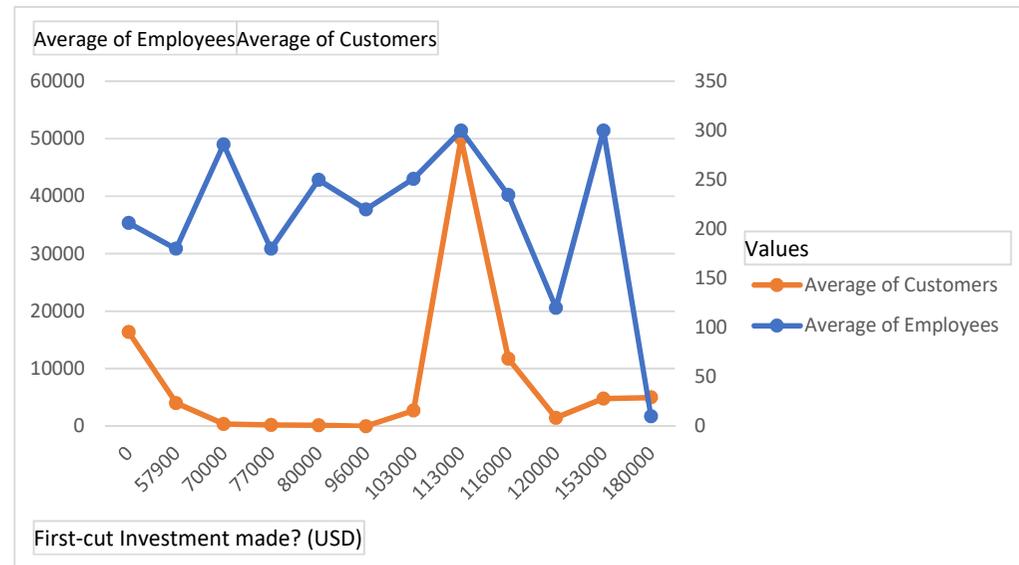

| Row Labels | Sum of ROI seen in first year? (USD) | Sum of ROI in 2017 (USD) |
|---|---|---|
| 2000 | $ - | $ - |
| 2001 | $ - | $ - |
| 2003 | $ 1,50,000 | $ 3,50,000 |
| 2004 | $ - | $ - |
| 2005 | $ - | $ - |
| 2006 | $ 1,13,000 | $ 13,56,000 |
| 2008 | $ 57,900 | $ 5,79,000 |
| 2009 | $ 1,53,000 | $ 13,77,000 |
| 2010 | $ 2,30,000 | $ 18,37,000 |
| 2011 | $ 96,000 | $ 6,72,000 |
| 2013 | $ 1,16,000 | $ 5,80,000 |
| 2015 | $ 1,90,000 | $ 5,70,000 |
| 2016 | $ 77,000 | $ 1,54,000 |
| 2017 | $ 50,000 | $ - |
| **Grand Total** | **$ 12,32,900** | **$ 74,75,000** |

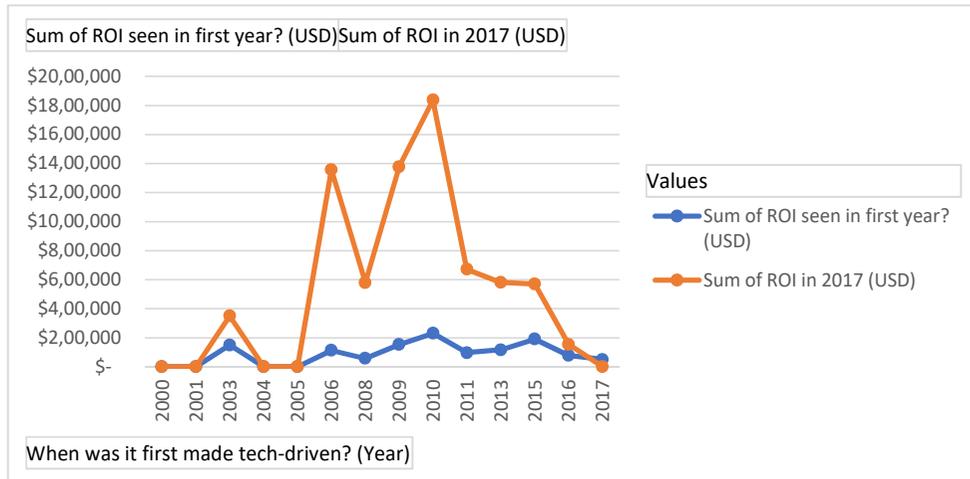

When was it first made tech-driven? (Year)